\documentclass[article]{JHEP3}
\usepackage{amsmath,epsfig}
\usepackage{amssymb,amsfonts}
\usepackage{epsfig}
\graphicspath{{images/}}

\relax
\renewcommand{\theequation}{\arabic{section}.\arabic{equation}}

\DeclareMathOperator{\End}{\text{End}}

\newcommand{\Real}{\mathbb{R}}

\newcommand{\Integer}{\mathbb{Z}}

\newcommand{\mat}{\begin{pmatrix}}
\newcommand{\tam}{\end{pmatrix}}
\newcommand{\smat}{\left(\begin{smallmatrix}}
\newcommand{\stam}{\end{smallmatrix}\right)}
\newcommand{\mc}{\mathcal}

\newcommand{\affH}{{\hat{H}_4}}
\newcommand{\nH}{{H_4}}

\newcommand{\aH}{{H_4/U(1)}}
\newcommand{\U}{{U(1)}}

\def\be{\begin{equation}}
\def\ee{\end{equation}}
\def\ba{\begin{eqnarray}}
\def\ea{\end{eqnarray}}

\def\tr{\text{tr}}

\def\g{\mathfrak{g}}
\def\h{\mathfrak{h}}
\def\ag{\mathfrak{\hat{g}}}
\def\ah{\mathfrak{\hat{h}}}

\def\ap{\a {\rm '}}

\def\F{\Phi}

\def\e{\epsilon}
\def\h{\eta}

\def\z{\zeta}
\def\m{\mu}
\def\n{\nu}
\def\r{\rho}
\def\s{\sigma}
\def\th{\theta}
\def\t{\tau}

\def\f{\varphi}
\def\a{\alpha}
\def\b{\beta}
\def\d{\delta}
\def\l{\lambda}
\def\g{\gamma}
\def\G{\Gamma}
\def\D{\Delta}
\def\w{\omega}

\def\la{\langle}
\def\ra{\rangle}

\def\nb{\nonumber}

\title{The abelian cosets of the Heisenberg group}
\author{Giuseppe D'Appollonio${}^{a,b}$ and Thomas Quella${}^c$ \\
${}^a$ Department of Mathematics, King's College London\\
The Strand, London WC2R 2LS, United Kingdom \vspace{0.2cm} \\
${}^b$ Dipartimento di Fisica dell'Universit\`a di Cagliari, INFN Sezione di Cagliari\\ 
Cittadella Universitaria 09042 Monserrato, Italy  \vspace{0.2cm} \\ 
${}^c$ Institute for Theoretical Physics, University of
Amsterdam\\ Valckenierstraat 65, 1018 XE Amsterdam, The Netherlands  }

\preprint{ITFA-2007-42 \\ KCL-MTH-07-13 }

\abstract{
In this paper we study the abelian cosets of the
$H_4$ WZW model. They coincide or are related to 
several interesting three-dimensional backgrounds 
such as the Melvin model, the conical point-particle 
space-times and the null orbifold. 
We perform a detailed CFT analysis of all the models
and compute the coset characters as well as some typical 
three-point couplings of coset primaries.}

\keywords{Conformal and W symmetry, Conformal Field Models in String Theory, Space-time Singularities}

\begin{document}

\maketitle 

\section{Introduction}

One of the principal aims of quantum gravity is the study and possible
resolution of space-time singularities. String theory is a very prominent candidate 
for a theory of quantum gravity and it provides several tools to deal with 
quantum corrections. However even in the context of string theory
the analysis of singular or time-dependent backgrounds 
is still a very subtle and difficult problem, both from a conceptual and 
a technical point of view. One of the best established approaches is to 
investigate the properties of backgrounds whose non-linear $\s$-models
are exact conformal field theories, since they represent
exact solutions to the
string equations of motion to all orders in the string tension
$\alpha'$. 

A first class of models is represented by the Lorentzian orbifolds 
of flat space \cite{hs}. 
Despite their apparent simplicity, the dynamics of these models 
and the nature of their singularities is still poorly understood 
\cite{cc,lms,pol,berkooz,null}. A second class of models is provided by
gravitational waves \cite{gwhis}. In this case the presence of a null isometry
allows to prove that the $\s$-model
is conformally invariant and to study some of its features in 
the light-cone gauge, as done for instance in \cite{prt}. Sometimes it is 
also possible to identify the chiral algebra of the underlying CFT and 
to solve the model in a covariant way. The 
first example of this type was discovered by Nappi and Witten \cite{nw} who
showed that the WZW model based on the Heisenberg group $H_4$ coincides with
the $\s$-model of the maximally symmetric gravitational wave in four dimensions. 
The exact solution of this model was given in \cite{dk, dk2}. 

The $H_4$ WZW model is also a particular example of a third large class
of exact curved backgrounds of string theory, the WZW and coset models based on non-compact groups
\cite{dpl,Gawedzki:1991yu,wbh}. For several years 
the lack of a proper understanding of the representation theory
of the corresponding affine algebras   
and the technical difficulties associated with the computation of the
correlation functions of a non-compact CFT, allowed only 
a semi-classical analysis 
of the geometries of these models, 
based on the gauged WZW Lagrangian \cite{coset}. 
Two prominent examples of this kind of constructions
are the two-dimensional black hole
\cite{wbh} 
and the cosmological Nappi-Witten space-time \cite{Nappi:1992kv}.
The situation changed during the past few years, when 
the structure of the $SL(2,\Real)$ WZW model and of the 
$H_3^+$ model were finally clarified
\cite{Teschner:1997ft,Maldacena:2000hw,Maldacena:2001km}
and the CFTs of some of their cosets were
analysed in more detail \cite{Maldacena:2000kv, Elitzur:2002rt}.
Although all these models can only be regarded as toy models for realistic
singularities, they provide a very convenient framework 
to explore the behaviour of strings in situations that, even if   
still unfamiliar, are likely to display new and interesting
physical effects.

In the present paper we initiate a systematic 
study of the cosets of the Heisenberg group $H_4$. 
There are several reasons to single out these models.
First of all the corresponding non-linear $\s$-models 
describe the propagation of strings in interesting curved space-times.
They were first considered in \cite{kk,sfetsos,ao,tse-sfet}
but at that time, as in the case of $SL(2,\mathbb{R})$ and its cosets, 
only a semi-classical analysis was possible.
We can now exploit the exact solution of the $H_4$ WZW model
\cite{dk,dk2} to provide a complete analysis of the underlying CFTs and in
particular to determine the operator content and the couplings of
all the models. 
Moreover
the Heisenberg group is non-compact and non-semisimple. 
Consequently, the structure of its cosets is remarkably rich
and is worth investigating in order to gain a better understanding
of the representation theory of this kind of chiral algebra.
This is particularly
evident in the case of the diagonal cosets \cite{ao,dq2} but also 
for the simpler abelian cosets discussed in this paper.

The abelian cosets of the Heisenberg group
provide a unifying framework for the study of several 
time-dependent or singular 
three-dimensional space-times.
They fall into two main classes
corresponding to two inequivalent $u(1)$ subalgebras.
The models in the first class \cite{sfetsos} 
have either Lorentzian or Euclidean signature. 
In the first case they are T-dual to the conical space-time
generated by a point source in three dimensions
\cite{djt1} while in the second case they coincide 
with the Melvin model \cite{Melvin}.
The models in the second class \cite{kk}
have a null isometry and describe  a   
gravitational wave with an infinite sequence
of singularities. It is interesting to note that 
these two classes, which  
arise from inequivalent gaugings of $H_4$,
are related by a singular limit. In terms of
its action on the coordinates, this limit is a Penrose limit
that leads from the point-particle space-time to the gravitational 
wave. It is possible to obtain a third class of new models by
an asymmetric
construction that involves simultaneously the two inequivalent $u(1)$ subalgebras.
The resulting backgrounds can again be interpreted 
as limits of the conical space-times.

We also show that the 
second class of cosets, when the spatial direction transverse 
to the wave is compactified on a circle, reduces in a suitable limit to the null orbifold
\cite{lms} and therefore provides a new framework for the study of the singularity 
of this model and its possible resolutions. 
It is worth mentioning that the fact that the three-dimensional
geometries described by the abelian cosets of the Heisenberg group
are T-dual to orbifolds of flat space is not surprising. In 
the free-field realization of the $H_4$ current algebra \cite{kk}, 
the primary fields are essentially twist fields. From this point of view
the affine algebra simply provides a very efficient tool for organising the 
spectrum of the model and for computing its correlation functions \cite{dk}. 

Exploiting the underlying chiral algebra, we determine the operator content
of all these models. This provides additional evidence
for the equivalence of the abelian $H_4$ cosets with the
three-dimensional backgrounds discussed above.
As an example we show in detail how the partition function
of the Melvin model \cite{tse-mel0,tse-mel,dudas,melvin} can be 
expressed in terms of coset characters. 
Moreover we show that the abelian coset that describes the three-dimensional
wave is an ordinary CFT. Hence the logarithmic CFT that in \cite{slog}
was conjectured to arise from a 
contraction of the $SU(2)$ parafermions does not seem to be directly
related to this geometric background.

Finally the knowledge of 
the three- and four-point amplitudes of the $H_4$ model \cite{dk} makes in principle also the
dynamics of these backgrounds accessible to a detailed investigation, at least at tree level
in the string coupling $g_s$. We explain how to evaluate a generic
amplitude for the cosets and then compute explicitly some
three-point couplings, leaving a more complete discussion 
and the analysis of four-point amplitudes to future work.

The plan of the paper is as follows. In Section~\ref{ml1} we
introduce our conventions and review the geometry of the Nappi-Witten gravitational
wave. 
In Section~\ref{ml2} we classify the 
possible $U(1)$ subgroups of $H_4$ and discuss the resulting abelian cosets
emphasizing their relations with the Melvin model, the conical space-times and the 
null orbifold. In Section~\ref{ml3} we 
determine the coset characters and the operator content of our models.
In Section~\ref{ml4} we explain how to evaluate 
the correlation functions and present
the calculation of some three-point couplings.
An Appendix summarizes our conventions and useful information
regarding the representation theory of the affine Heisenberg algebra.

\section{The maximally symmetric plane wave}
\label{ml1}

The WZW model based on the Heisenberg group $H_4$ describes the propagation of 
a string in a homogeneous Lorentzian space which is, as first noticed in \cite{nw}, 
a gravitational wave supported by a non-trivial flux $H_{\m\n\r}$ of the two-form field $B_{\m\n}$.
Gravitational wave metrics are usually presented in 
Brinkmann or Rosen coordinates and in our case we have 
\begin{align}
& ds^2_{b} = 2dudv_b - \frac{\r^2}{4}du^2+d\r^2 + \r^2d\chi^2 \ , & &  
H_{u \r \chi} = \ - \r  \ , &
\label{brink}  \\
& ds^2_r = 2dudv_r +\sin^2(u/2) (dx_1^2+dx_2^2) \label{rosen}\ , & &
H_{u x_1 x_2} = - \sin^2 \frac{u}{2}  \ . &
\end{align}
For the analysis of the coset models two additional coordinate systems
will be particularly useful
\begin{align}
& ds^2_{1} = 2dudv_1 - (a_1da_2-a_2da_1)du + da_1^2 + da_2^2   \ ,   & & H_{ua_1a_2} = 1 \ , &\label{nw1}\\
& ds^2_{2} = 2dudv_2 + dx^2 + dy^2 + 2 \cos u dxdy   \ ,  & & H_{uxy} = \sin{u} \ . &\label{nw2}
\end{align}
One can pass from one coordinate system to the other using the transformations 
\begin{align}
& a_1 = x + y \cos u \ , & & 
a_2 = y \sin u \ , & &
v_1 = v_2 + \frac{xy}{2} \sin u \ ,  \\
& \z = e^{iu/2}(a_1-ia_2) \ , & & 
v_b = v_1 \ , & &  \\
&  \z = xe^{iu/2} + ye^{-iu/2} \ ,  & & 
v_b = v_2 + \frac{xy}{2}\sin u \ , & &  \\
&  \z = (x_1+i x_2) \sin{u/2} \ , & & 
v_b = v_r - \frac{x_1^2+x_2^2}{8}\sin{u} \ , & & 
\end{align}
where $\z = \r e^{i \chi}$. The action of the non-linear $\s$-model 
associated with this gravitational wave 
coincides with the action of the WZW model of the $H_4$ group.
The previous four coordinate systems correspond to the 
following four parametrisations of the group elements
\begin{align} & g_{1} =
e^{a_1P_1 + a_2P_2}e^{uJ+v_1K} \ ,  & &
g_{2} =
e^{xP_1} e^{uJ}e^{yP_1+v_2K} \ ,  \\
& g_{b} =
e^{\frac{u}{2}J} e^{\frac{\bar{\z}}{2}P^-+
\frac{\z}{2}P^+}e^{\frac{u}{2}J+v_bK} \ , &  &
g_{r} = e^{\frac{x_2P_1+x_1P_2}{2}} e^{u J} e^{-\frac{x_2P_1+x_1P_2}{2}+v_rK}\ .
\end{align}
Here the  anti-hermitian operators $P_1$, $P_2$, $J$ and $K$ are the
four generators of the Lie algebra of the Heisenberg group 
and satisfy the commutation relations
\be
[P_1,P_2] = K \ , \hspace{1cm} [J,P_1] = P_2 \ , \hspace{1cm}
[J,P_2] = - P_1 \ . \ee
Further details on $H_4$ and its representations are collected in appendix \ref{rt}.

The $H_4$ WZW model was completely solved in \cite{dk,dk2}, where the three- and four-point 
correlation functions 
were computed for bulk and boundary vertex operators with maximally
symmetric boundary conditions. In this paper we perform a similar analysis for the abelian
cosets of $H_4$ \cite{kk,sfetsos}. Before studying the underlying coset chiral algebras in detail,
we discuss their geometric interpretation which relies on the
Lagrangian formulation of the gauged WZW models \cite{coset}.

\section{The abelian cosets of the Heisenberg group}
\label{ml2}

To construct an abelian coset of the maximally symmetric plane wave,
we have first to single out a specific direction 
in the Lie algebra of $H_4$. This direction, by the non-compactness of 
the group manifold, can 
be either time-like, space-like
or null. Moreover the subgroups
for the left and right action of $U(1)$ on $H_4$ can be chosen
almost independently. 
Let us consider two embeddings $\e_{L,R}$ of a $u(1)$ algebra with generator 
$Q$ into the $H_4$ Lie algebra. 
If the two embeddings satisfy 
the constraint 
\be
\bigl\la \e_L(Q) \e_L(Q) \bigr\ra = \bigl\la \e_R(Q) \e_R(Q) \bigr\ra \ , \label{cc} \ee
where the brackets denote a fixed invariant bilinear form on the $H_4$ Lie algebra, 
the gauge transformation
\be
g \mapsto e^{s \, \e_L(Q)} g e^{-s \, \e_R(Q)} \ , \ee
is anomaly free and leads to a consistent gauged 
WZW model \cite{coset}. 
From the form of the gauge transformation it is clear that the coset
model only depends on the conjugacy classes of $\e_L(Q)$ and of $\e_R(Q)$,
since even independent left and right inner automorphisms simply result
in a reparametrisation of the group element. 
Consequently there are two main classes of abelian cosets of the $H_4$ WZW
model. 

The first class consists of all the models for which
both $\e_L(Q)$ and $\e_R(Q)$ contain the generator $J$. 
In this case the $u(1)$ subalgebra can 
be brought to a standard form by conjugation and it
is generated by a linear combination 
of $J$ and $K$
\be
\e_L(Q) = \a J + \b K \ , \hspace{3cm}  \e_R(Q) = \bar \a
 J + \bar \b K \ . \label{a1}
\ee The embeddings are specified by 
four real parameters subject to the relation $\a \b = \bar \a
\bar \b$, so that the gauging is anomaly free. Without loss of generality
we can choose $2 \a \b = \xi$ where $\xi = \pm 1$, with the 
two values of $\xi$ corresponding to different choices for the signature 
of the embedded $U(1)$.
As we will see in the next subsection, the coset models with $\xi = 1$ 
coincide with the 
conical space-times associated with a particle in three dimensions 
while the coset models with $\xi = -1$ lead to the Melvin model. 
When one of the parameter in Eq.~\eqref{a1} vanishes, the embedded $U(1)$ is null.
The corresponding coset models are extremely simple since they are
flat and two-dimensional, although their algebraic construction presents
a few subtleties \cite{nullg}. 
The Lagrangian description of the models with 
$\a = \bar \a$ and $\a = - \bar \a$, which
correspond respectively to the vector 
and axial gauging of the $u(1)$ current, 
was first considered in \cite{sfetsos}. 
Let us note that the vector and the axial gauging are not
related by an inner automorphism.

The second class consists of all the models for which
neither $\e_L(Q)$ nor $\e_R(Q)$ contain the generator $J$. 
Without loss of generality we can choose 
\be \e_L(Q) = P_1 , \hspace{3cm}  \e_R(Q) = - P_1  \ . \label{b1}
\ee
In this case the axial and vector gauging are related by an inner
automorphism. The geometric data of this model, first obtained
in \cite{kk},
describe a gravitational wave in three dimensions
with a non-trivial dilaton field. The metric has a sequence of singularities
close to which it reduces to the null orbifold considered
in \cite{lms}. The agreement is precise if
one compactifies the spatial direction transverse to the wave. 

There is also a third class of models defined by the following embedding
\be
 \e_L(Q) = P_1 \ , \hspace{3cm}  \e_R(Q) =  -\bar \a J - \frac{1}{2 \bar \a} K  \ . \label{c1}
\ee
These asymmetric cosets describe a class of singular three-dimensional space-times
that can be interpreted as limits of the conical space-times.
In the rest of this paper we will focus mainly on the first 
two classes of models and describe in detail the $\a J + \b K$ cosets and the
$P_1$ cosets.

\subsection{The Melvin and the conical space-times}

Let $g \in H_4$ and consider the following gauge 
transformation \be g \mapsto e^{ s (\a J + \b K)} g e^{- s  (\bar \a
 J + \bar \b K)} \ . \ee 
In the coordinate system (\ref{nw1}) 
we have
\be (u,v,r,\f) \mapsto (u+ s (\a-\bar
\a),v+ s (\b - \bar \b), r \, , \f + s \a) \ , \ee where  
$a_1+i a_2 = r e^{i \f}$. 
There are two convenient gauge choices. If $\a \ne
\bar \a$ we can fix  completely the gauge freedom setting $u = 0$.
Alternatively we can first set $\f = 0$ and then take into account the residual
discrete identifications
\be
(u, v)  \sim \left (u + 2 \pi k \, \frac{\a - \bar \a}{\a} \ , v + 2 \pi k \, \frac{\b - \bar \b}{\a} 
\right ) \ , 
\hspace{1cm} k \in \mathbb{Z} \ .
\ee
It is convenient to define the parameter $\h$
\be
\h = \frac{\a - \bar \a}{\a \bar \a} \ , \ee
where we are assuming that both $\a$ and $\bar \a$ are non-zero. 
The metric, dilaton and two-form fields of the model with $\xi = 1$ are
\be
ds^2 = dr^2 + \frac{\h^2 r^2}{\h^2 - r^2}  d\f^2 - \frac{4 }{\h^2 - r^2} dv^2
\ ,   \hspace{0.5cm}
B = \frac{2 \h^2}{\h^2 - r^2} d\f \wedge dv \ , \hspace{0.5cm} \F
= \ln (\h^2 - r^2) \ . \label{mel+}\ee
This is a three-dimensional Lorentzian space with a naked singularity
and closed time-like curves in the
region $r^2 > \h^2$ where the time direction is the periodic coordinate $\f$.
The model with $\xi = -1$  corresponds to the following smooth Euclidean space
\be ds^2
= dr^2 + \frac{\h^2 r^2 }{\h^2 + r^2}d\f^2+\frac{4 }{\h^2 +
r^2}dv^2
\ , \hspace{0.5cm}
B = \frac{2 \h^2}{\h^2 + r^2} d\f \wedge dv \ , \hspace{0.5cm} \F =
\ln (\h^2 + r^2) \ . \label{mel-} \ee
When $\a = \bar \a$ 
the geometric data of the coset
are \be ds^2 = - \xi dx^2 + dr^2 + \frac{4 dv^2}{r^2} \
, \hspace{1cm} \F = \ln (r^2) \ , \label{dualsp} \ee 
with $x = u/\a$. This metric can also be derived from 
Eq.~\eqref{mel+} and Eq.~\eqref{mel-} setting $x = \h \f$ and 
taking the limit $\h \rightarrow 0$.

The space-times in Eq.~\eqref{mel+} and Eq.~\eqref{mel-} are T-dual 
to freely acting orbifolds of flat space. 
The model with $\xi = -1$ is in fact easily recognized as the Melvin 
model \cite{Melvin, gmel1, gmel2, tse-mel0, tse-mel}. Let us start with a flat three-dimensional
Euclidean space \be ds^2 = dx^2 + dr^2 + r^2 d \th^2 \ , \ee 
and identify the points related by the transformation
$(x, \th) \mapsto (x + 2 \pi R, \th + 2 \pi \g)$ with $R \in \mathbb{R}$ and $\g \in [0,1)$. 
We then define a new coordinate $\f = \th - \frac{\g}{R}
x$ and perform a T-duality along the $x$ direction. The resulting 
background coincides with $(\ref{mel-})$ if we set
$\h = \frac{R}{\g}$ and $2 v \g = R \tilde{x}$ ,
where $\tilde{x}$ is the coordinate T-dual to $x$. As such the coordinate $v$
has radius $R_v = \frac{1}{\g}$ so that the generic Melvin background coincides
with an abelian coset of the $H_4$ model
with a compact $v$ direction. The existence of this relation between the $H_4$
model and the Melvin space-time was first noticed
in \cite{tse-mel}.

The model with $\xi = 1$ is T-dual to the conical space-time generated by a point mass
in three dimensions \cite{djt1}. This is equivalent to a three-dimensional Minkowski
space $ds^2 = -dt^2 + dr^2 + r^2 d\th^2$ where 
the points related by the transformation 
$( t, \th ) \mapsto (t - 2 \pi R, \th + 2 \pi \g )$ are identified. The  
mass and angular momentum of the source are
\be
m = \frac{1-\g}{4 \pi G_N} \ , \hspace{1cm} L = \frac{R}{4 G_N}  \ , \ee
where $G_N$ is the Newton constant.
In order to compare these conical space-times and the models in Eq.~\eqref{mel-} 
we follow exactly the same steps as before,   
introducing a new coordinate $\f = \th + \frac{\g}{R} t$ 
and then performing a T-duality along the $t$ direction.  
The relations between the parameters of the two models are
$\h = \frac{R}{\g}$, $\h \tilde t = 2v$ and $v \sim v - 2\pi/\g$,
so that
the background generated by a point particle with generic mass and angular
momentum coincides with an abelian coset of the $H_4$ model
with a compact $v$ direction of radius $R_v = 1/\g$.

Also the model in Eq.~\eqref{dualsp} is T-dual to a flat three-dimensional
space. To see this one first writes the metric of the two-plane in 
polar coordinates and then performs a T-duality along the angular
direction. This model was discussed in \cite{telias} as a toy model
to study T-duality transformations in non-compact spaces.

As a final observation, let us stress that 
the abelian cosets of $H_4$ can be related to suitable
limits of   
the two-dimensional charged black-hole
and the three-dimensional black string \cite{cbh, hh1, gq}
or more generally to abelian cosets of models given by 
the product of $U(1)$, $SU(2)$ and $SL(2, \mathbb{R})$
WZW models.
This is due to the fact that 
the affine algebra based on the Heisenberg group is a contraction of the 
product of $\widehat{SU}(2)$
and a time-like free boson or of $\widehat{SL}(2, \mathbb{R})$ and
a space-like free boson \cite{ors}.

\subsection{The gravitational wave and the null orbifold}

Let us consider now the second class of abelian cosets and gauge
the $u(1)$ subalgebra generated by $P_1$.
Let $g \in H_4$ and consider the gauge 
transformation \be g \mapsto e^{ s P_1} g e^{\mp s P_1} \ . \ee 
The two choices of sign correspond respectively to the vector and axial
gauging of the $U(1)$ current and are related by an inner automorphism.
In the coordinate system (\ref{nw2}) the transformation is simply
$x \mapsto x + s$, $y \mapsto y \mp s$, 
so that we can fix the gauge freedom setting $x = \! y$ for the vector model
and $x = - y$ for the axial 
model. The background fields of the cosets are 
easily obtained and describe a
three-dimensional gravitational
wave supported by a dilaton \cite{kk}. For the axial gauging we have 
\be ds^2 = 2 du dv +
4 \tan^2 \frac{u}{2} \, dx^2 \ ,
\hspace{2cm} \F=  \ln \cos^2 \frac{u}{2}  \ ,  \label{3dgwt} \ee
where the result is automatically expressed
in Rosen coordinates. The coordinate transformation
\be v = V + \frac{X^2}{2\sin u} \ ,
\hspace{2cm} x = \frac{X}{2} \cot \frac{u}{2}  \ , \label{r2b} \ee brings the
metric into Brinkmann form 
\be ds^2
= 2 du dV + \frac{X^2}{2\cos^2 \frac{u}{2}}du^2 + dX^2 \ , \hspace{2cm} \F= \ln \cos^2 \frac{u}{2}  \ . \ee
The background fields for the vector gauging
are related to the previous ones by
$ u \rightarrow \pi - u$, $v \rightarrow - v$. 
The metric is singular at the points $u_n = n \pi$.
The  $R_{uxux}$ component
of the Riemann tensor of the axial model diverges 
at the points $u_{2n+1}$ where the model is also
strongly coupled. Around these points a better description of the 
background geometry is provided by the T-dual vector model.

We show now that in a neighbourhood of $u = 0$ the metric and the wave functions
of the $P_1$ cosets reduce to the corresponding quantities of the null
orbifold considered in \cite{lms}.
In order to take the limit in a controlled way we 
first rescale the coordinates
\be
(u, v, x) \rightarrow (\z u, v/ \z, x/ \z) , \ee
and then send the parameter $\z$ to zero.
In this limit the dilaton of the axial model tends to 
a constant while the metric becomes
\be
ds^2 = 2 du dv + u^2 dx^2 \ , \hspace{3cm} ds^2 = 2 du dV + dX^2 \ .  
\label{lno1}
\ee
The null orbifold is a Lorentzian orbifold of flat space,
a class of models introduced in \cite{hs}. It owes its name to the fact that  
the orbifold group is generated by a null Lorentz transformation, namely 
a linear combination of a boost and a rotation. We start from
$ds^2 = 2 du dV + dX^2$ and identify the points related by
\be
\left ( X, V \right )  \sim \left ( X + 2 \pi  \n u \ , V - 2 \pi \n X - 2 \pi^2 \n^2  u 
\right ) \ , \ee
where the value of the parameter $\n$ can be set to one by a boost. 
The change of coordinates
\be
X = u x \ , \hspace{1cm} V = v - \frac{u}{2} x^2 \ , \ee
brings the metric into the form $(\ref{lno1})$ with the identification
$ x \sim x + 2 \pi \n$. Therefore in order to have precise agreement with 
the geometry of the $P_1$ coset we must compactify the coordinate $x$ 
of the latter on a circle
of radius $R$. If we denote by $i \l$ and $i \bar \l$ the eigenvalues of $P_1$,
the axial model contains only winding states with $\l + \bar \l = 0$ while the
vector model contains only momentum states with $\l - \bar \l = 0$. 
A discrete spectrum containing both winding and momentum modes 
requires a compact $x$ direction. In this
case the operators that survive in the spectrum have
\be
\l = \frac{n}{R} + \frac{w R}{2} \ , \hspace{2cm}
\bar \l = \frac{n}{R} - \frac{w R}{2} \ . \ee
It is this compactified coset model with radius $R = 1/\z$ that precisely matches the 
null orbifold in the limit $ \z \rightarrow 0$. Note that unlike
the parameter $\n$ of the null orbifold, the radius $R$ labels different models 
and cannot be set to one by a boost. 

To illustrate this correspondence further let us show that 
the wave functions of the states of the coset become in the limit the
semi-classical wave functions of the null orbifold. 
The wave function of a state in $H_4$ with $K = ip$, $P_1 = i \l$, $\bar P_1 = 
i \bar \l$ and Casimir $C$ is 
\be
\psi_{p,C,\l,\bar \l} \, = \, \frac{1}{\sqrt{|\sin u |}} e^{ - i p v - i \frac{C}{p} u - i \l x + i \bar \l y
- \frac{i}{4 p}\left [ (\l + \bar \l)^2 \cot \frac{u}{2}  - (\l - \bar \l)^2 \tan \frac{u}{2} \right ]}
\label{nw2wf} \ . \ee
In order to reduce this wave functions to the coset we set $y = -x$. 
In Brinkmann coordinates the result is 
\be
\psi_{p, C,\l, \bar \l} \, = \, \frac{1}{\sqrt{|\sin u |}} e^{ - i p V - i \frac{C}{p} u 
- i \frac{p}{4} \tan \frac{u}{2} \left [ X^2 - \frac{(\l - \bar \l)^2}{p^2} \right ] 
- i \frac{p}{4} \cot \frac{u}{2} \left [ X^2 + \frac{\l + \bar \l}{p} \right ]^2} 
 \ . \ee
We now rescale the coordinates and take  the limit $\z \rightarrow 0$ 
keeping $\tilde p = p/ \z$ finite in the limit. 
Up to a shift in the value of the Casimir $C$ the wave function becomes 
\be
\psi_{\tilde p, C, n, w} \, \sim \, \frac{1}{\sqrt{|u|}} e^{ - i \tilde p V - i 
\frac{C}{\tilde p} u 
- i \frac{\tilde p}{2u} \left ( X + \frac{2n}{p} \right )^2 + \frac{i}{4 \tilde p} \frac{w^2}{24} u^3} 
 \ . \ee
Following the notation of \cite{lms}, the previous expression 
coincides with the wave function of a state of the null orbifold 
with winding number $w$, momentum $J = n$ and light-cone momentum 
$p^+ = \tilde p/2$. Moreover, $u = - 2 x^+$ and $V = x^-/2$.
All the states whose wave function is given by Eq.~\eqref{nw2wf} have $p \ne 0$.
The wave functions of states with $p = 0$ are distributions localised at the points
$u_n$ and can be identified with the states that live on the
singular plane crossing the tip of the two cones of the null orbifold.

It is interesting to note that this background is also the Penrose limit 
of the point particle space-times,
that is of the $\a J + \b K$ cosets with $\xi = 1$. 
If we set
\be
r = \h \sin \frac{u}{2} \ , \hspace{1cm} \f = \frac{x}{\h}
\ , \hspace{1cm} v = \frac{\h^2}{8}(u + \sin u ) - t \ , \ee
the new coordinates parametrize the region $r < \h$. 
In the limit $\h \rightarrow \infty$ we magnify the neighbourhood of a null
geodesic and 
the space-time reduces to the three-dimensional gravitational wave
discussed in this section. Although
the $U(1)$ subgroup generated by $P_1$ and the one
generated by $\a J + \b K$  with $\xi = 1$ are not equivalent 
the relation
\be  \a J + \frac{1}{2\a} K =  e^{-\frac{P_2}{\a}}(\a J + P_1)e^{\frac{P_2}{\a}} \ , 
\label{plc}
\ee
implies that they are connected by the singular limit
$\a \rightarrow 0$.
This limit is the algebraic equivalent of
the Penrose limit and it is very similar 
to the relation between the elliptic orbifold
$ \mathbb{R} \times \mathbb{C}/\mathbb{Z}_N$ and the null orbifold 
discussed in \cite{lms, null}.

As a final remark let us mention that 
this gravitational wave can also be obtained as the Penrose limit
of the parafermions times a time-like free boson. The background fields  of the coset
$SU(2)_k/U(1)_k \times \mathbb{R}$ are
\be
ds^2 = -kdt^2 + k [d\th^2+\tan^2 \th d\f^2] \ ,  \hspace{1cm} 
\F = \ln \cos^2 \th \ , 
\ee
and if we set
\be
u = t + \th \ , \hspace{1cm} \frac{2v}{k} = -t + \th \ , \hspace{1cm}
\frac{2x}{\sqrt{k}} = \f \ , \ee
and consider the limit $k \rightarrow \infty$ we obtain $(\ref{3dgwt})$.
From the point of view of the theory on the world-sheet, the Penrose limit
is a contraction of the chiral algebra. The contractions of  
$SU(2)_k \times U(1)$ to $H_4$ and of $SL(2,\mathbb{R})_k \times SU(2)_k$ to 
$H_6$ were studied in detail in \cite{dk, bdkz}. 
It would be interesting to study also the $P_1$ coset 
as a contraction of the chiral algebra of the parafermions and a time-like boson.
This point of view was adopted in \cite{slog} where
the authors defined a contracted chiral algebra starting with
a superposition of parafermionic and $u(1)$ operators 
whose conformal dimensions agree only in the limit. When one 
expands the OPEs in a power series in the contraction parameter, 
the subleading difference in the conformal weights yields additional terms 
that are typical for a logarithmic CFT. 
This observation led the authors of \cite{slog} to conjecture that 
the resulting theory is a logarithmic CFT with $c = 3$. 

It seems worthwhile to point out that the limit so defined cannot
correspond to the geometric background described in this paper. In
fact, the abelian $H_4$ coset is certainly not a logarithmic CFT
since the dilatation operator $L_0$ is completely diagonalisable on
the full state space. This can most easily be understood on the level of
the
minisuperspace theory where $L_0$ reduces to the Laplacian acting on a
suitable space of functions. For the coset this space coincides with the
space of functions on the Heisenberg group which are invariant under
the action of the subgroup. Also, the coset Laplacian agrees with that
of the Heisenberg group up to a contribution proportional to
$P_1^2$. Both of these operators are completely diagonalisable,
thus proving our assertion.
Presumably 
this property persists in
the full CFT and the correlation functions of the primary vertex
operators satisfy the standard conformal Ward identities.
A non-diagonalisable term in $L_0$ could appear if 
one considers the supersymmetric version of this coset or even the supersymmetric $H_4$ 
WZW model itself. This is due to the fact that the world-sheet fermions 
transform in the adjoint representations of the Heisenberg group
which is not fully decomposable. A similar non-diagonalisable
term in $L_0$ was discussed in the context of the 
null orbifold of the RNS superstring \cite{null}. 

\subsection{The third class of models}

Here we briefly describe the models that result from gauging
the symmetry 
\be
g \mapsto e^{s \, P_1}   \, g \, e^{ s \left ( \bar \a J + \frac{K}{2 \bar \a} \right )} \ ,
\ee
or in the coordinate system \eqref{nw1}
\be
(u, v, a_1, a_2) \mapsto (u + \bar \a s, a_1 + s, a_2, v + s a_2/2 + s/2 \bar \a) \ . \ee
We choose the gauge $a_1 = 0$ and set $a_2 = x$. The background fields are
\be
ds^2 = \frac{1}{1+ \bar \a x} \left [ 2 du dv - \frac{x}{\bar  \a} du^2 \right ] + dx^2 \ , 
\hspace{0.6cm} B = \frac{\bar  \a x}{1 + \bar  \a x } \, du \wedge dv \ , 
\hspace{0.6cm}  \F = \log (1 +\bar  \a x) \ . 
\label{tcm}
\ee
This is a curved space-time with Ricci scalar 
$R = - \frac{7 \bar  \a^2}{2(1 + \bar  \a x)^2}$ and 
a singularity at $ x = - 1/\bar  \a$. It follows from Eq.~\eqref{plc}
that this background is another limit of the conical space-time
\eqref{mel+}. We first rewrite the metric \eqref{mel+}
with  $\h = \frac{1}{\a}+\frac{1}{\bar \a}$
in Cartesian  coordinates $a_1 = r \cos \f$, $a_2 = r \sin \f$ 
and set
\be a_1 = \frac{u}{\bar \a} \ , \hspace{1cm}
a_2 = x - \frac{1}{ \a} \ , \hspace{1cm}
v_1 = - v + \frac{u}{2 \bar \a} \left(x + \h \right ) \ . 
\ee
The limit $ \a  \rightarrow 0$ gives the background fields in Eq.~\eqref{tcm}.

\section{The coset characters}
\label{ml3}

The knowledge of the exact CFT underlying the 
curved backgrounds discussed in the previous section allows us to compute
the spectrum of the vertex operators and their correlation functions. 
When these CFTs are used to build a consistent string background,
we can determine both the exact spectrum of perturbative string excitations
and their interactions. The operator content
of the abelian cosets of $H_4$ can be derived from the spectrum of
the $H_4$ model \cite{kk,eb,dk}.
The spectrum of
a coset conformal field theory $G/H$ is given by a collection
of usually irreducible representations of the coset chiral algebra. 
Any representation $\m$ of the affine Lie algebra $\ag$ associated to
$G$ can in fact be decomposed into representations
of the affine subalgebra $\ah\subset\ag$ belonging to the subgroup
$H$. The resulting branching spaces give rise to 
representations of the coset chiral algebra, the commutant of $\ah$
in the universal enveloping algebra $\mc{U}(\ag)$.
In the following we shall describe the decomposition of the $\affH$ representations
with respect to the two abelian subalgebras 
that have been used to construct the backgrounds discussed in the previous section.
For a review of the representation theory of the affine Lie algebra 
$\affH$  we refer the reader to Appendix~A.

\subsection{The operator content of the first class of models}

The characters of the first class of models can be obtained from
the $\affH$ characters in a straightforward way. 
This is due to the fact that $J$ and $K$ are the generators of the 
standard Cartan
subalgebra of $H_4$. The character formulas for the coset representations 
resemble closely those appearing in
the context of the two-dimensional black hole \cite{Pak, Rib}.
We will decompose the $\affH$ characters with respect to the
$\hat{u}(1)$-subalgebra defined by the injection
\begin{equation}
  \epsilon(Q)\ =\ \alpha J + \beta K \ \ ,
\end{equation}
with $2\alpha\beta=\xi \in\{\pm1\}$ depending on whether we are
gauging a space-like or a time-like isometry.
It follows from this condition
that the map $\epsilon$ is an embedding of the
full current algebra 
\begin{equation}
  Q(z)\,Q(w)\ =\ \frac{\xi}{(z-w)^2}\ \ .
\end{equation}
  Via $\epsilon$ any $\affH$-representation $\mu$ can be interpreted
  as a reducible $\hat{u}(1)$-representation and the irreducible
  constituents $\chi_{[\mu, \, b]}^\aH(q)$ can be determined by writing 
\ba
  \label{eq:DecoIdeaJK}
  \tr_{\mu}\Bigl[q^{L_0-\frac{c}{24}}\,z^{-i\epsilon(Q_0)}\Bigr]
  \ &=&\ \tr_{\mu}\Bigl[q^{L_0-\frac{c}{24}}\,z^{-i\alpha J_0}\,z^{-i\beta K_0}\Bigr]
  \ =\ \chi_{\mu}^\nH(q,z^\alpha,z^\beta) \nb \\
  \ &=&\ \sum_b\chi_{[\mu, \, b]}^\aH(q)\,\chi_b^\U(q,z)\ \ ,
\ea
where in the last expression we introduced the 
standard $\hat{u}(1)$ characters of charge $b$
\begin{equation}
  \chi_b^\U(q,z)\ =\ \frac{q^{\frac{\xi}{2}b^2}\,z^b}{\eta(q)}\ \ .
\end{equation}
For the continuous series of representations, Eq.~\eqref{eq:DecoIdeaJK}  
leads immediately to the coset characters
\begin{equation}
  \chi_{[(0,\s,j)_{-\w}, \, b^0_{n}]}^\aH(q)
  \ =\ \frac{q^{h_{(0,\s,j)} + \w (j + n) - \frac{\xi}{2}
(b^0_{n})^2 }}{\eta(q)^3}\ \ , \hspace{1cm} \w \in \mathbb{Z}  \ , 
\label{cmc}
\ee where the charge $b^0_{n}$ is given by
\be
b^0_{n} = \alpha(j+n) + \frac{\xi}{2\a}\w  \ . 
\label{chzero}
\end{equation}
The conformal dimension of the corresponding coset
representation is 
\begin{equation}
  h_{[(0,\s,j)_{-\omega}, b^0_{n}]}^\aH
  \ =\ h_{(0,\s,j)} + \w(j+n) -\frac{\xi}{2} (b^0_{n})^2 \ \ .
\end{equation}
In order to obtain the characters of the discrete series we 
first expand 
$\chi_{(\pm,p,j)_\omega}^\nH(q,z^\alpha,z^{\beta})$
according to Eq.~\eqref{eq:AffineCharactersTwo}  
and then regroup the $z$-dependence. The
resulting expression reads
\begin{equation}
  \chi_{[(\pm,p,j)_{\mp \w},  \, b^\pm_{n}]}^\aH(q)
  \ =\ \sum_{m=0}^\infty(-1)^m\frac{q^{h_{(\pm,p,j)}+\frac{m}{2}(m + 2n+1)
\pm \omega(j \mp n)-\frac{\xi}{2} (b^\pm_{n})^2}}{\eta(q)^3}\ \ , 
\label{cmd} \ee
where the $U(1)$ charges $b^\pm_{n}$ are \be
b^\pm_{n} = \alpha(j \mp n) \pm \frac{\xi}{2\a}(p + \w) \ . 
\label{chpm}
\end{equation}
  If $n\geq0$ the minimal contribution to the $q$-series comes
  from $m=0$. On the contrary for $n<0$  the first few terms in the series cancel
  and the minimum of the exponent
  is reached for a positive half-integer value of $m$.
  Altogether we end up with the conformal dimensions
\begin{equation}
  h_{[(\pm,p,j)_{\mp \w},  \, b^\pm_{n}]}^\aH
  \ =\ \begin{cases}
         h_{(\pm,p,j)} \pm \w(j \mp n) -\frac{\xi}{2}  \, (b^\pm_{n})^2 &,
\ n \geq0\\[2mm]
        h_{(\pm,p,j)} \pm \w(j \mp n) -\frac{\xi}{2}  \, (b^\pm_{n})^2 - n&,
\  n<0\ \ .
       \end{cases}
\end{equation}

The states of the WZW model survive the
coset projection precisely when $Q=\bar{Q}$. 
When $\a\ne \bar \a$ it is convenient to label the states 
in the decomposition of a discrete representation 
with the values of $p$, $\w$, $n$ and $\bar n$ and then use the 
previous constraint to fix $j$
\be
j = \pm \xi \frac{p+\w}{2 \a \bar \a} \pm  \frac{\a n - \bar \a \bar n}{\a - \bar \a} \ .
\label{cons1}
\ee
In a similar way we label the states in
the decomposition of a continuous representation 
with the values of $\s$, $\w$ and of $L = n - \bar n$. This set of quantum numbers
identifies the state uniquely, since the values of  
$j \in [0,1)$ and $n + \bar n$ are given by the constraint
\be j + n = \xi \frac{\w}{2 \a \bar \a}  - \frac{\bar \a L}{\a - \bar \a} \ . 
\label{cons2} \ee
When $\a = \bar \a$ the constraint $Q=\bar{Q}$ reduces to $n = \bar n$
and therefore we can specify $p$, $j$, $\w$ and $n$ for the 
states that appear in 
the decomposition of the discrete representations $(\pm,p,j)_{\mp \w}$,
and $\s$, $\w$, $j$ and $n$ for the  states that appear in 
the decomposition of the continuous representations $(0,\s,j)_{-\w}$.

The spectrum of this class of coset models can be neatly summarised by the
torus partition function. When $\a \ne \bar \a$ we obtains
\ba
{\cal Z}_{H_4/U(1)} &=& \sum^\infty_{\w = 0} \int_0^1 dp \sum_{n, \bar n \in \mathbb{Z}}
\left ( \left | \chi_{[(+,p,j)_{- \w},  \, b^+_{n}]}^\aH(q) \right |^2 + 
\left | \chi_{[(-,p,j)_{\w},  \, b^-_{n}]}^\aH(q) \right |^2 \right ) \nb \\
&+&
 \sum^\infty_{\w \in \mathbb{Z}} \int_0^\infty ds s 
\, \int_{0}^{2 \pi} d \b \, \sum_{L \in \mathbb{Z}}
 \left | \chi_{[(0,\s,j)_{-\w},  \, b^0_{n}]}^\aH(q) \right |^2 \ ,
\ea
where the value of the quantum numbers that do not appear in the sums or in the
integrals is assumed to be fixed by $(\ref{cons1})$ or $(\ref{cons2})$.
We now discuss the case $\xi = -1$ in more detail in order to show the precise
relation between the abelian $H_4$ coset and the Melvin model. 
Using Eq.~\eqref{cmc} and~\eqref{cmd}, the contribution of the discrete representations can be written as
\ba {\cal Z}^{{\rm discrete}}_{H_4/U(1)} &=& 
\sum_{\w\in \mathbb{Z}} \int_0^1 dp  \sum_{n, \bar n \in \mathbb{Z}}
q^{ \frac{1}{2}\left [ \frac{\h}{2}(p+\w) - \frac{n - \bar n}{\h} \right ]^2 + \frac{p}{2}(1- p) + n p }
\bar q^{ \frac{1}{2}\left [ \frac{\h}{2}(p+\w) - \frac{n - \bar n}{\h} \right ]^2 + \frac{p}{2}(1- p) + \bar n p }
\nb \\ && \frac{1}{[\h(\t)\h(\bar \t)]^3} \sum^\infty_{m, \bar m = 0}(-1)^{m + \bar m}
q^{\frac{m}{2}(m + 2n +1)}\bar q^{\frac{\bar m}{2}(\bar m + 2 \bar n +1)}
\ , \label{jkpf}
\ea
and the contribution of the continuous representations as
\be {\cal Z}^{{\rm continuous}}_{H_4/U(1)} = 
\frac{1}{[\h(\t)\h(\bar \t)]^3}
\sum_{\w, L \in \mathbb{Z}}  \int_{0}^{2 \pi} d \b \, \int_0^\infty ds \, s \, 
q^{ \frac{s^2}{2} + \frac{1}{2}\left [ \frac{\h}{2} \w - \frac{L}{\h} \right ]^2}
\bar q^{ \frac{s^2}{2} + \frac{1}{2}\left [ \frac{\h}{2} \w + \frac{L}{\h} \right ]^2}
\ . \label{jkpfc}
\ee
We want to compare Eq.~\eqref{jkpf} with the contribution of the twisted sectors to
the partition function of the
Melvin model. 
The partition function of the Melvin model has been discussed in several papers 
\cite{tse-mel, dudas, melvin} since it provides
an interesting example of a non-compact orbifold. The contribution of the twisted
sectors can be written, setting $\ap = 2$,
\be
{\cal Z}^{{\rm twisted}}_{{\rm Melvin}} = \sum_{s, t \in \mathbb{Z}, s \ne 0} \int_{- \infty}^\infty
\frac{d l}{2 \pi} \, q^{\frac{1}{2}(l + sR/2)^2} \bar q^{\frac{1}{2}(l - sR/2)^2}
e^{2 \pi i (l R) t} \left | \vartheta \left [ \begin{matrix} 1/2 + s \g \cr 
1/2 + t \g  \end{matrix} \right ](0 | \t) \right |^{-2} \ ,
\ee
where $\vartheta[a,b](z|\t)$ is the Jacobi $\vartheta$-function. 
Using the identity $(\ref{eq:CharIdOne})$ we can write this partition function in the 
following form which allows a simple comparison with the 
contribution of the discrete representations in the coset partition function
\ba
{\cal Z}^{{\rm twisted}}_{{\rm Melvin}} &=&
\sum_{s, k \in \mathbb{Z}, s \ne 0}  \sum_{n, \bar n \in \mathbb{Z}}
q^{ \frac{1}{2}\left [ \frac{\g(n - \bar n) + k}{R} - \frac{sR}{2} \right ]^2 + \frac{\{\g s \}}{2}(1-\{\g s \})}
\bar q^{ \frac{1}{2}\left [ \frac{\g(n - \bar n) + k}{R} + \frac{sR}{2} \right ]^2 + \frac{\{\g s \}}{2}(1-\{\g s \})}  
\nb \\ &&  \frac{q^{n\{\g s \}} \, \bar q^{\bar  n\{\g s \}}}
{[\h(\t)\h(\bar \t)]^3} \sum^\infty_{m, \bar m = 0}(-1)^{m + \bar m}
q^{\frac{m}{2}(m + 2n +1)}\bar q^{\frac{\bar m}{2}(\bar m + 2 \bar n +1)}
\ , \label{mel2}
\ea
where $\{ a \}$ stands for the fractional part of the real number $a$.
As explained in the previous section the $H_4/U(1)$ model corresponds to the limit
\be
 \frac{R}{\g} \rightarrow \h \ , \hspace{1cm} \g \rightarrow 0 \ , \ee
of the Melvin model. In this small radius limit the momentum modes labeled by $k$
decouple and the discrete sum over the twisted sectors
becomes an integral in a continuous variable that can be
identified with $p+\w$. 

When $\g$ is an irrational number, the states in a twisted sector 
with $\g s$ very close to an integer are almost delocalised.
The limit of these almost delocalised twisted states are   
the spectral flowed continuous representations of the coset model. 

In order to reproduce the partition function of the Melvin model
with a finite radius $R = \h \g$, we have to compactify the 
$v$ direction of the $H_4$ model on a circle of radius $R_v = 1/\g$. As a result 
the quantum numbers $p + \w$ and $\z \equiv j - \bar j$ can only be multiples of 
the compactification radius and its inverse
\be p + \w = s \g \ , \hspace{2cm}
\z = \frac{k}{\g} \ , \ee
and the partition function 
\ba {\cal Z}^{{\rm discrete}}_{H_4/U(1)} &=& 
\sum_{ s, k \in \mathbb{Z}}   \sum_{n, \bar n \in \mathbb{Z}}
q^{ \frac{1}{2}\left [ \frac{\h}{2}(p+\w) - \frac{n - \bar n + \z}{\h} \right ]^2 + \frac{p}{2}(1- p) + n p }
\bar q^{ \frac{1}{2}\left [ \frac{\h}{2}(p+\w) - \frac{n - \bar n - \z}{\h} \right ]^2 + \frac{p}{2}(1- p) + \bar n p }
\nb \\ && \frac{1}{[\h(\t)\h(\bar \t)]^3} \sum^\infty_{m, \bar m = 0}(-1)^{m + \bar m}
q^{\frac{m}{2}(m + 2n +1)}\bar q^{\frac{\bar m}{2}(\bar m + 2 \bar n +1)}
\ , \label{jkpfcompact}
\ea
agrees precisely with $(\ref{mel2})$.

The partition function of the model with $\a = \bar \a$ and
of the conical space-times can be discussed
along similar lines. 

\subsection{The operator content of the second class of models}

The irreducible representations of the Heisenberg group are classified by the
values of the central element $K$ and of the Casimir operator $C$.
Until now we have labeled the states in a given representation
using a discrete label $n$ related to their $J$ eigenvalues.
More precisely
the states in the representation
$(\pm,p,j)$ have $J$-eigenvalues $j \mp n$, $n \in \mathbb{N}$ and are denoted
by $|\pm, p, j; n\ra$ 
while the 
states in the representation
$(0,\s,j)$ have $J$-eigenvalues $j + m$, $m \in \mathbb{Z}$ and are
denoted by $|0, \s, j; m\ra$.
Instead of diagonalising the generator
$J$, we can also diagonalise $P_1$. If we do so, the states
in each irreducible representation carry a continuous label $\lambda$
since the spectrum of the generator  $P_1$ is continuous. 
To study the characters of the $P_1$ coset it is natural to choose this
continuous basis. 
The relation between the basis of $J$ and $P_1$ eigenstates is quite simple
\ba
| \pm, p, j; \l \ra &=& \frac{e^{-\frac{\l^2}{2p}}}{(\pi p)^{1/4}}\sum_{n=0}^\infty
\frac{H_n(\l/\sqrt{p})}{2^{\frac{n}{2}}\sqrt{n!}} |\pm, p, j; n\ra \ , \nb \\
| \pm, p, j; n \ra &=& \int_{-\infty}^\infty d \l \frac{e^{-\frac{\l^2}{2p}}}{(\pi p)^{1/4}}
\frac{H_n(\l/\sqrt{p})}{2^{\frac{n}{2}}\sqrt{n!}} |\pm, p, j; \l\ra \ ,
\label{changebasis}
\ea
where $\l \in \mathbb{R}$ and the functions $H_n(t)$ are the Hermite polynomials.
It is easy to verify that $P_1 | \l, p, j \ra = i \l | \l, p, j \ra$.
Using Mehler's formula we can evaluate the following trace
on a representation of the horizontal subalgebra
\be
\tr_{(\pm,p,j)} \left (z^{-i J} x^{-iP_1} \right ) =  \int d \l \, x^\l \t_{(\pm,p)}(z,\l) 
\chi_{(\pm,p,j)}(z)\ .
\ee
where
\be
\t_{(-,p)}(z,\l) = 
\frac{1}{\sqrt{\pi p}} \,
\sqrt{\frac{1 + z}{1 - z}} e^{-\frac{\l^2}{p}\frac{1-z}{1+z}}
 \ ,
\label{mehler}
\ee
and $\t_{(+,p)}(z,\l) = \t_{(-,p)}(1/z,\l)$.
Similar relations hold for the continuous representations.
In this case we can construct the states
\be
|0, \s, j;\th \ra = \sum_{m \in \mathbb{Z}} e^{i m \th} |0, \s, j;m \ra \ ,
\hspace{1cm} \th \in [0,2\pi) \ ,
\ee
which are eigenstates of $P_1$ with eigenvalue $\l = s \cos (\b + \th)$.
The evaluation of the trace gives in this case
\be
\tr_{(0,\s,j)} \left (z^{-i J} x^{-iP_1} \right ) =   \frac{1}{\pi}
\int_{-s}^s \frac{d \l \, x^\l}{\sqrt{s^2-\l^2}} 
 \sum_{n\in\Integer} z^{j+n}  \ .
\ee
In order to decompose the $\affH$ characters with respect to the
subalgebra generated by $P_1$ we first 
study the decomposition of the affine representations $\hat{\mu}$ 
with respect to the horizontal subalgebra. This is easily done using
once more  eq.\ \eqref{eq:CharIdOne}. 
For the discrete representations we obtain 
\begin{equation}
  \label{eq:AffineCharactersThree}
  \chi_{(\pm,p,j)_{\mp \w}}^\nH(q,z)
  \ =\ \sum_{n\in\Integer}\frac{z^{j \mp n}}{1 - z^{\mp 1}} q^{h_{(\pm,p,j)} \pm \w (j \mp n)}
 \sum_{m=1}^\infty(-1)^{m+1}\frac{q^{\frac{m}{2}(m+2n-1)}}{\eta(q)^4} (q^\w-q^m)\
\ . \ee
Once the character is written in this form, we can easily read the
multiplicity of the representation $(\pm,p,j\mp n)$
of the horizontal subalgebra at each level of the affine representation. 
We then use this decomposition
and the trace in $(\ref{mehler})$ to evaluate the following modified $\hat{H}_4$-character
\ba
  && \tilde \chi_{(\pm,p,j)_{\mp \w}}^\nH(q,z,x)
  \ \equiv \ \tr_{(\pm,p,j)_{\mp \w}}\bigl(q^{L_0-\frac{c}{24}}\, z^{-i J_0} \,x^{-i(P_1)_0}\bigr)  
   = \ \sum_{n \in \mathbb{Z}} \int d\l \,  x^\l \, \t_{(\pm,p + \w)}(z,\l) \nb \\
&&\qquad\qquad \chi_{(\pm, p, j\mp n)}(z)
 q^{h_{(\pm,p,j)} \pm \w (j \mp n)}
 \sum_{m=1}^\infty(-1)^{m+1}\frac{q^{\frac{m}{2}(m+2n-1)}}{\eta(q)^4} (q^\w-q^m)
 \ . \nb \ea
Finally, in order to extract the coset characters we set $z = 1$ and remove 
the contribution of the affine $u(1)$ character. The function $\t_{(\pm,p + \w)}(z,\l)$
is divergent for $z=1$ but this divergence can be absorbed 
in the normalization of the states of the non-compact $U(1)$. 
The result is 
\be
 \chi_{[(\pm,p,j)_{\mp \w},\l]}^\aH(q) 
=   \sum_{n \in \mathbb{Z}}  
q^{h_{(\pm,p,j)} \pm \w (j \mp n)- \frac{\l^2}{2}}
\sum_{m=1}^\infty(-1)^{m+1}\frac{q^{\frac{m}{2}(m+2n-1)}(q^\w-q^m)}{\eta(q)^3}
 \ . \ee
When $\w = 0$ the character can be written in the simple form 
\be 
 \chi_{[(\pm,p,j),\l]}^\aH(q) =  
\frac{q^{h_{(\pm,p,j)} - \frac{\l^2}{2}}}
{\eta(q)^3}
 \ . \ee
Following the same 
strategy as outlined above we obtain from the decomposition
of the continuous representations the characters
\be
 \chi_{[(0,\s,j)_{-\w},\l]}^\aH(q,z,u)
  = \sum_{n\in\Integer}\frac{q^{h_{(0,\s,j)}- \frac{\l^2}{2} + \w(j+n)}}{\eta(q)^3}  z^{j+n} 
\ ,
\ee
where $\l$ can assume the values $\l = s \cos \th$ with $0 \le \th < 2 \pi$.
The conformal dimensions of the coset representations are therefore
\be
h_{[(\pm,p,j)_{\mp \w};\l]} = \pm (p + \w) j + \frac{p}{2} (1 - p) - \frac{\l^2}{2} \ , 
\ee with
$\w \in \mathbb{N}$, $(\l, j) \in \mathbb{R}^2$, $p \in (0,1)$ 
and 
\be
h_{[(0,\s,j)_{- \w};\l]} = \w j + \frac{s^2 - \l^2}{2} \ ,  \ee
with $\w \in \mathbb{Z}$, $s \in \mathbb{R}_+$, $\l \in [-s, s]$, 
$j \in [0,1)$.
The spectral flowed representations, as it is evident from their characters, 
contain states with arbitrarily negative values of $L_0$.

\subsection{The operator content of the third class of  models}

In order to describe the spectrum of
the third class of models we only have to identify 
the states that survive the coset
projection,
since we already know the decomposition of the $H_4$ characters
with respect to the abelian subalgebras generated by $P_1$ and by
$\a J + \b K$.
The states of the coset belong to the
representations described by the characters 
\be \chi^\aH_{[(\pm, p, j)_{\mp \w}, \, \l_\pm]} \, \, \bar \chi^\aH_{[(\pm, p, j)_{\mp \w}, \, b_n^\pm]}
\ ,  \hspace{1cm}
\ee
with $p \in (0,1)$, $\w \in \mathbb{N}$, $j \in \mathbb{R}$, $n \in \mathbb{Z}$
and $\l_\pm = - b_n^{\pm}$ or by the characters 
\be
 \chi^\aH_{[(0, \s, j)_{\w}, \, \l_0]} \, \, \bar \chi^\aH_{[(0, \s, j)_{\w}, \, b_0^\pm]} \ , 
\ee
with $s \in \mathbb{R}_+$, $(\w, \, n) \in \mathbb{Z}$, $j \in [0,1)$ and $\l_0 = - b_n^0$. Here the
charges $b_n^0$ and $b_n^\pm$ are as defined in Eq.~\eqref{chzero} and \eqref{chpm}.


\section{Interactions}
\label{ml4}

The correlation functions of a coset model $G/H$ can be
expressed in terms of the correlation functions of the 
$G$ and $H$ WZW models (see for instance \cite{gepner}). When applied to the vertex operators,
the decomposition of 
the $G$ representations into a sum of products of coset and $H$ representations
gives
\be
\F_g^G(z) = \sum_{h} \F_{[g,h]}^{G/H}(z) \, \F_h^H(z) \ ,
\ee
where  $ \F_g^G$,  $ \F_h^H$ and $ \F_{[g,h]}^{G/H}$ 
are respectively primary fields of the two WZW models and of the coset model. 
To simplify the notation we are displaying only the chiral part of the 
vertex operators.
A general $n$-point correlator can then be written in a factorized form
\be
\left \la \prod_{i=1}^n \F_{g_i}^G(z_i) \right \ra
= \sum_{h_i} \left \la   \prod_{i=1}^n \F_{[g_i,h_i]}^{G/H}(z_i) \right \ra 
\left \la  \prod_{i=1}^n \F_{h_i}^H(z_i)  \right  \ra \ ,
\ee
from which one can extract the correlators of the coset model
and in particular the two- and three-point couplings.
Since the three- and four-point correlation functions of the $H_4$ WZW model
are known \cite{dk}, we can use these results to evaluate some typical correlators 
for the abelian cosets discussed in the previous sections. 
We restrict our attention to correlators that do not involve spectral flowed states.

Let us denote a primary field of $H_4$ as $\F^\nH_{(q,n,\bar n)}$, where
$q$ stands for all the quantum numbers necessary to specify
the representation of $\affH$ and $n$, $\bar n$ keep track of the $J$ and $\bar J$ eigenvalues
of the field. With this notation $(q,n,\bar n)$ identifies the state with eigenvalues
$j \mp n$ and $j \mp \bar n$ if $q = (\pm,p,j)$ and with eigenvalues
$j + n$ and $j + \bar n$ if $q = (0,\s,j)$. 
The correlation functions assume a simpler form if one introduces
charge variables to handle simultaneously all the fields that belong
to the same representation \cite{fz}.
For fields in a
discrete representation we use a complex variable $x$. The monomials
\be
\b_{p,n}(x) = \frac{(\sqrt{p} \, x)^n}{\sqrt{n!}} \ , \hspace{1cm} n \ge 0 \ , \label{cvd} \ee
provide a complete orthonormal basis for the space of analytic functions
$f(x)$ with measure
\be
d \s_p =  \frac{d^2 x}{\pi} p \, e^{ - p x^* x} \ , \ee
where $x^*$ is the complex conjugate of $x$. We also introduce
a complex charge variable $\bar x \in \mathbb{C}$ for the anti-chiral fields.
For fields in a continuous representation we consider the phases
$\b_{0,n}(\th) = e^{i n \th}$ with $\th \in [0, 2\pi)$, $n \in \mathbb{Z}$ and use the measure 
$d \s_0 = d \th/ 2\pi$. 
We can now collect all the primary fields that belong to the same representation
in a single field that depends both on the worldsheet and the charge variables
\be
\F^{\nH}_q(z, \bar z, x, \bar x) = \sum_{n, \bar n} \F^\nH_{(q,n,\bar n)}(z, \bar z) \b_{q,n}(x)
\b_{q,\bar n}(\bar x) \ , \ee
with $n$, $\bar n \in \mathbb{N}$ for the discrete representations and 
$n$, $\bar n \in \mathbb{Z}$ for the continuous representations.
The three-point couplings of these primary fields are given by
\be
\left \la \prod_{i=1}^3 \F^\nH_{q_i}(x_i, \bar x_i) \right \ra = {\cal K}_{q_1,q_2,q_3}(x_1,x_2,x_3)
\, {\cal K}_{q_1,q_2,q_3}(\bar x_1, \bar x_2, \bar x_3) \, {\cal C}_{q_1,q_2,q_3} \ . \label{gtpc} \ee
Here and in the following we leave understood the standard dependence 
on the insertion points $z_i$ of the vertex operators, which is completely 
fixed by the Ward identities of global conformal invariance.
The explicit form of the functions ${\cal K}_{q_1,q_2,q_3}$  and ${\cal C}_{q_1,q_2,q_3}$ 
can be found in \cite{dk,bdkz,dk2}.\footnote{The $(\pm,p,j)$ 
representations in the conventions of \cite{dk,bdkz,dk2} 
correspond to the $(\mp, p, j)$ representations in the conventions of the present paper.}

Let us consider first some examples of three-point couplings for the $\a J + \b K$ cosets.
In this case we need the couplings between the fields $\F^\nH_{(q,n, \bar n)}$ that 
can be easily extracted from the couplings in Eq.~\eqref{gtpc} using the orthonormality of the 
functions $\b_{q,n}$ and the quantities
\be 
{\cal K}^{n_1,n_2,n_3}_{q_1,q_2,q_3} \equiv \prod_{i=1}^3 \int
d \s_{q_i} \b^*_{q_i,n_i}(x_i) \, {\cal K}_{q_1,q_2,q_3}(x_1,x_2,x_3) \ . 
\ee
Consider the coupling between the coset fields $[(-,p_1,j_1),b^-_{n_1},b^-_{\bar n_1}]$, 
$[(-,p_2,j_2),b^-_{n_2},b^-_{\bar n_2}]$ and $[(+,p_3,j_3),b^+_{n_3},b^+_{\bar n_3}]$, where
the charges $b^\pm_n$ are as defined in Eq.~\eqref{chzero} and Eq.~\eqref{chpm}. 
In the following  we assume for simplicity that all the labels $n_i$ and $\bar n_i$
are non-negative, otherwise one needs the couplings of descendant fields of
the $H_4$ WZW model.
When the $q$ labels take the previous values, the $H_4$ coupling is non-zero
only when $p_3 = p_1 + p_2$ and $\Delta \equiv -(j_1+j_2+j_3) \in \mathbb{N}$. In this
case we have  
\be {\cal K}_{q_1, q_2, q_3} = e^{-x_3(p_1x_1+p_2x_2)}(x_2-x_1)^\Delta \ ,
\hspace{1cm} 
{\cal C}_{q_1, q_2, q_3} = 
\frac{1}{\Delta !}\left [ \frac{\g(p_3)}{\g(p_1)\g(p_2)} \right ]^\Delta \ ,
\ee
where $\g(x) = \frac{\G(x)}{\G(1-x)}$. 
The three-point coupling of the coset fields is then given by 
\begin{multline}
  \qquad\qquad
\left \la \F^{H_4/U(1)}_{[(-,p_1,j_1),b^-_{n_1},b^-_{\bar n_1}]}
\F^{H_4/U(1)}_{[(-,p_2,j_2),b^-_{n_2},b^-_{\bar n_2}]}
 \F^{H_4/U(1)}_{[(+,p_3,j_3),b^+_{n_3},b^+_{\bar n_3}]} \right \ra\\
  =  
\frac{1}{\D!}\left [ \frac{\g(p_1+p_2)}{\g(p_1)\g(p_2)} \right ]^\Delta
\, {\cal K}^{n_1,n_2,n_3}_{q_1, q_2, q_3} \, 
 {\cal K}^{\bar n_1,\bar n_2,\bar n_3}_{q_1, q_2, q_3}
 \ ,\qquad\qquad
\end{multline}
where $\Delta = n_1 + n_2 - n_3 = \bar n_1 + \bar n_2 - \bar n_3$ and 
\be {\cal K}^{n_1,n_2,n_3}_{q_1, q_2, q_3} = 
\sqrt{\frac{n_2!}{n_1!n_3!}}
\frac{\D! \, (-1)^{n_1+n_3}}{(n_2-n_3)!} \frac{p_2^{n_3}}{p_1^{n_1/2} p_2^{n_2/2} p_3^{n_3/2}  }
F(-n_1,-n_3;n_2-n_3+1;-p_1/p_2) \ ,  \ee
when $n_2 \ge n_3$ and 
\be
{\cal K}^{n_1,n_2,n_3}_{q_1, q_2, q_3} = 
\sqrt{\frac{n_1!n_3!}{n_2!}}
\frac{ (-1)^{n_1+n_2}}{(n_3-n_2)!} \frac{p_1^{n_3-n_2}}{p_1^{n_1/2} p_2^{n_2/2} p_3^{n_3/2}  }
F(- \Delta,-n_2;n_3-n_2+1;-p_1/p_2) \ ,  
\ee
when $n_3 \ge n_2$. Here $F(a,b;c;t)$ is the hypergeometric function.
As a second example we consider 
the coupling between the coset fields $[(-,p_1,j_1),b^-_{n_1},b^-_{\bar n_1}]$, 
$[(+,p_2,j_2),b^+_{n_2},b^+_{\bar n_2}]$ and $[(0,\s,j_3),b^0_{n_3},b^0_{\bar n_3}]$
and for simplicity we assume again that the labels $n_i$ and $\bar n_i$, $i = 1, 2$
are non-negative. In this case
we have a non-vanishing $H_4$ coupling only if $p_1 = p_2 \equiv p$ and
$\Delta \equiv -(j_1+j_2+j_3) \in \mathbb{Z}$. Moreover 
\be {\cal K}_{q_1, q_2, q_3} = 
e^{-p x_1x_2 - \frac{1}{\sqrt{2}}\left(\s \, x_2x_3+ \s^* \frac{x_1}{x_3}\right)}x_3^\Delta \ ,
\hspace{1cm} 
{\cal C}_{q_1, q_2, q_3} = e^{\frac{s^2}{2} \Sigma(p)} \ ,
\ee
where
\be
\Sigma(p) = \psi(p)+\psi(1-p)-2\psi(1) \ , \hspace{1cm}  
\psi(x) = \frac{{\rm d} {\rm ln} \G(x)}{dx} \ .
\ee
When these conditions are satisfied the three-point coupling in the coset model 
can be written as
\be
\left \la \F^{H_4/U(1)}_{[(-,p,j_1),b^-_{n_1},b^-_{\bar n_1}]}
\F^{H_4/U(1)}_{[(+,p,j_2),b^+_{n_2},b^+_{\bar n_2}]}
 \F^{H_4/U(1)}_{[(0,\s,j_3),b^0_{n_3},b^0_{\bar n_3}]} \right \ra =  
e^{\frac{s^2}{2} \Sigma(p)}  \, {\cal K}^{n_1,n_2,n_3}_{q_1, q_2, q_3} \, \, 
 {\cal K}^{\bar n_1,\bar n_2,\bar n_3}_{q_1, q_2, q_3} \ ,
\ee
where $\Delta = n_1 - n_2 + n_3 = \bar n_1 - \bar n_2 + \bar n_3$ and  
\be  {\cal K}^{n_1,n_2,n_3}_{q_1, q_2, q_3} = 
\left ( \frac{s}{\sqrt{2p}} \right )^{n_1+n_2} 
\sum_{k=0}^{m}\frac{(-1)^{k+n_1+n_2}\sqrt{n_1!n_2!}}{k!(n_2-k)!(n_1-k)!}
\left ( \frac{2p}{s^2} \right )^{k} \ , \ee
with $m = {\rm min}(n_1,n_2)$.
Following the same procedure one can compute various other couplings among
coset fields.

Let us consider now the second class of models. 
The best way to compute the couplings of the $P_1$ coset is to first rewrite
the $H_4$ couplings in a basis of $P_1$ and $\bar P_1$ eigenstates.
To this end we apply the change of basis in Eq.~\eqref{changebasis} to the states
\be
|\pm, p, j, x, \bar x \ra = \sum_{n, \bar n = 0}^\infty 
\b_{p,n}(x) \b_{p,\bar n}(\bar x) |\pm, p, j, n, \bar n \ra  \ .
\ee
The result is 
\ba
|\pm, p, j, x, \bar x \ra &=&  \int_{- \infty}^{\infty} d\l 
\int_{- \infty}^{\infty} d\bar \l \, G_p(x,\l)  \, G_p(\bar x,\bar \l)
|\pm, p, j, \l, \bar \l \ra  \ , \nb \\
|\pm, p, j, \l , \bar \l \ra &=&  \int_{\mathbb{C}} d\s_p  \int_{\mathbb{C}} d \bar \s_p  
 \, G_p^*(x,\l)  \, G_p^*(\bar x,\bar \l)
|\pm, p, j, x, \bar x \ra  \ ,
\ea
where
\be
G_p(x,\l) = \frac{1}{(\pi p)^{1/4}} e^{-\frac{\l^2}{2p}+\sqrt{2}\l x
-\frac{p}{2}x^2} \ , \hspace{1cm} \l \in \mathbb{R} \ , x \in \mathbb{C} \ .
\ee
For the continuous representations we consider the states
\be
|0, \s, j, \th, \bar \th \ra = \sum_{n, \bar n \in \mathbb{Z}} 
\b_{0,n}(\th) \b_{0,\bar n}(\bar \th) |0, \s, j, n, \bar n \ra  \ ,
\ee
which are eigenstates of $P_1$ with eigenvalue $ \l = s \cos(\th + \bar \th + \b)$. 
Once the $H_4$ couplings have been expressed in a basis of $P_1$
eigenstates, the couplings of the coset fields can be easily obtained
by removing the contribution of the $U(1)$ vertex operators and setting
\be \l = \frac{n}{R} + \frac{mR}{2} \ , \hspace{1cm}
 \bar \l = \frac{n}{R} - \frac{mR}{2} \ , \ee
if we consider the general case of the compact coset with $x \sim x + 2 \pi R$.
The condition imposed on the states by the vector gauging ($\l = \bar \l$)
and the condition imposed on the states by the
axial gauging ($\l = - \bar \l$) arise respectively in
the limit $R \rightarrow \infty$ and $R \rightarrow 0$. 
This is of course what we expect, 
since the axial and the vector model are related by T-duality.
We present here just two examples of three-point couplings.
The first one is
\ba
&& \left \la \F^{H_4/U(1)}_{[(-,p_1,j_1),\l_1,\bar \l_1]}
\F^{H_4/U(1)}_{[(-,p_2,j_2),\l_2,\bar \l_2]}
 \F^{H_4/U(1)}_{[(+,p_3,j_3),\l_3,\bar \l_3]} \right \ra =   \\
&&  \frac{1}{\D!}\left [
\frac{\g(p_3)}{\g(p_1)\g(p_2)} \right ]^{\Delta+\frac{1}{2}} \left [
\frac{p_3}{2p_1p_2} \right ]^{\Delta}
e^{-\sum_{i=1}^3 \frac{\l_i^2+\bar \l_i^2}{2p_i}}  H_\D (\m ) H_\D ( \bar \m )
 \ , 
\ea 
where $p_3 = p_1 + p_2$, 
$\D \in \mathbb{N}$, $\l_1 + \l_2 + \l_3 = \bar \l_1 + \bar \l_2 + \bar \l_3 = 0$ and 
the argument $\m$ of the Hermite polynomial $H_\D$ is 
\be
\m \equiv \sqrt{\frac{\l_1^2}{p_1} +\frac{\l_2^2}{p_2} - \frac{\l_3^2}{p_3}} \ , 
\ee 
The quantity $\bar \m$ has exactly the same expression in terms of the barred 
quantities.
Our second example is
\ba
&&  \left \la \F^{H_4/U(1)}_{[(-,p,j_1),\l_1,\bar \l_1]}
\F^{H_4/U(1)}_{[(+,p,j_2),\l_2,\bar \l_2]}
 \F^{H_4/U(1)}_{[(0,\s,j_3),\l_3,\bar \l_3]} \right  \ra =  \\
&&
e^{-\frac{\l_1\l_2}{2p}-\frac{\l^2_3}{2p} \tan^2 \th 
+\frac{i \, \l_3(\l_1-la_2)}{p} \tan \th} 
e^{-\frac{\bar \l_1\bar\l_2}{2p}-\frac{\bar\l^2_3}{2p} \tan^2 \th
+ \frac{i \, \bar\l_3(\bar\l_1-\bar\l_2)}{p} \tan \th}
e^{\frac{s^2}{2}\Sigma(p)}
\ ,
\ea
where  $\D \in \mathbb{Z}$, $\l_1 + \l_2 + \l_3 = \bar \l_1 + \bar \l_2 + \bar \l_3 = 0$
and $\l_3 = s \cos (\th + \b)$, $\bar \l_3 = s \cos (\bar \th + \b)$.

The vertex operators of the third class of models can be written as the
product of a chiral vertex operators of the  $P_1$ coset and an
anti-chiral vertex operators of the $\a J + \b K$ coset.
The computation of their correlation functions can then be 
performed following the steps already described
in this section.

\section{Conclusions}
\label{ml5}

In this article we reported on progress in understanding various
space-times arising from gauging a one-dimensional subgroup of
the Heisenberg group $H_4$. 
All the models we obtained are
non-compact and many of them in fact exhibit time-dependence and an
interesting structure of curvature singularities. Building on the 
solution of the $H_4$ WZW model \cite{dk,dk2}, we have been
able to derive exact conformal field theory results such as
spectra and correlation functions.

The specific backgrounds we discussed came in three families. In order to
obtain the first class we gauged a subgroup generated by
a linear combination of the currents $J$ and $K$. This resulted in
models which are T-dual to freely acting orbifolds of flat space.
Depending on the signature we recovered either the Melvin model
\cite{Melvin} or the conical space-times that are generated by point
particles in three-dimensional gravity \cite{djt1}. For the
Melvin model we showed explicitly that the coset partition
function coincides with the partition functions of a freely
acting orbifold of flat space \cite{tse-mel0,tse-mel}. 

The second class of models was obtained by gauging the subgroup
generated by $P_1$. Vector and axial gauging turned out to be equivalent
in this case and led to a gravitational wave with a periodic array of
singularities. While these two gaugings led to a continuous spectrum
of either momentum or winding modes, we also considered a compactified
version of the model, with a discrete spectrum of momenta and windings.
We showed, both at the level of the geometry and of the 
semi-classical wave functions, that this compact model
reduces in a suitable limit to the null orbifold of Minkowski space \cite{lms}.
We also briefly discussed a third class of models obtained by an asymmetric
construction involving simultaneously the two inequivalent $u(1)$ subalgebras
of $H_4$.

Our work invites for generalizations in several directions. For a better 
understanding of the models and of 
their singularities one should study in detail their dynamics. 
The role played by the twisted sectors 
of the Lorentzian orbifolds has been the object of several papers
\cite{lms,berkooz,null}
and led to some proposals for the fate of their singularities \cite{null},
inspired by the analogy with previous work on tachyon condensation in non-compact Euclidean 
orbifolds \cite{Adams:2001sv}.
It would be interesting to  
study the amplitudes of the compact $P_1$ coset 
discussed in this paper and to compare them with the results obtained 
for the null orbifold \cite{lms}.
It would also be worth investigating the space of the possible relevant and marginal
deformations of these models, taking for instance as a starting point
the current-current deformations of the $H_4$ WZW model
discussed in \cite{dk}, since 
the cosets considered here are 
the boundary points of such deformations. 

Additional insights on the physical properties of these space-times
may be gained by the analysis of the dynamics of open strings. 
Although in the present work we restricted
ourselves to the closed string sector, 
the entire analysis could be
extended to cover open strings and D-branes. In fact 
the boundary CFTs associated with the maximally symmetric branes of the Heisenberg
group were solved in \cite{dk2} and therefore the corresponding spectra 
and correlation functions are known.
Together with the formalism developed for the description of D-branes 
in asymmetrically gauged WZW models \cite{cosetbranes}, these results 
should lead to a clear picture of open strings in the abelian $H_4$ cosets.

Finally, it remains to say that the insights and techniques gained and developed
here can also be
used to study more complicated, non-abelian cosets based on the
Heisenberg group. This concerns in particular its diagonal cosets
which have been shown to correspond to non-maximally symmetric
plane waves in four dimensions \cite{ao}. We shall
return to this class of models in a forthcoming
publication \cite{dq2}.

\enlargethispage{3mm}

\acknowledgments The authors are grateful to 
Sylvain Ribault and Volker Schomerus
for discussions.
G.D. acknowledges the support of the PPARC rolling grant 
PP/C507145/1. The research of T.Q. is funded by a Marie Curie Intra-European Fellowship,
contract number MEIF-CT-2007-041765.
We also acknowledge partial support from the EU Research Training Network 
{\it Superstring theory}, 
MRTN-CT-2004-512194.

\appendix

\vskip 10mm
 \renewcommand{\theequation}{\thesection.\arabic{equation}}

\section{The group $H_4$ and its representations}
\label{rt}

The Lie algebra of the group $H_4$ has four generators 
that satisfy the following commutation relations
\be
[P_1,P_2] = K \ , \hspace{1cm} [J,P_1] = P_2 \ , \hspace{1cm}
[J,P_2] = - P_1 \ , \ee
or, in terms of the raising and lowering operators $P^\pm = P_1 \pm i P_2$,
\be
[P^+ , P^-] = -2i K  \ , \hspace{1cm} [J,P^{\pm}] = \mp i P^{\pm} \ .
\ee
All generators are anti-hermitian and the
invariant bilinear form is \be 
\la P^+ , P^- \ra = - 2 \ , \hspace{1cm} \la J, K \ra = - 1 \ . \ee 
The corresponding current algebra is 
\ba
P^+(z)\,P^-(w) &=& - \frac{2}{(z-w)^2}-\frac{2i K}{z-w} \ , 
\hspace{1cm} J(z)K(w) = - \frac{1}{(z-w)^2} \ , \nb \\
J(z)\,P^\pm(w) &=& \mp\frac{i P^\pm}{z-w} \ , 
\ea
or, in terms of the modes,
\be
[P^+_n, P^-_m] = -2n \d_{n+m,0} - 2i K_{n+m} \ , \hspace{0.4cm}
[J_n, P^{\pm}_m] = \mp i P^{\pm}_{n+m} \ , \hspace{0.4cm}
[J_n, K_m] = - n \d_{n+m,0} \ .
\label{affh4}
\ee
The energy momentum tensor 
\begin{equation}
T\ \ =\ - \frac{1}{4}\left[ P^+P^- + P^-P^+ \right ] - JK + \frac{K^2}{2} \ , 
\end{equation}
generates a Virasoro algebra with central charge $c=4$.

There are three classes of irreducible unitary representations of the group 
$H_4$ \cite{kk}:
two families of so-called discrete series representations $(\pm,p,j)$ 
(with $p > 0$ and $j\in\Real$) and one family of continuous
series representations $(0,\s,j)$ (with $\s = s e^{i \b}$, $s\geq 0$, $0 \le \b < 2 \pi$ 
and $j\in[0,1)$).
All these representations are infinite dimensional and their
weight content is clearly displayed by their characters 
$\chi(z,u)=\tr[z^{-iJ}u^{-iK}]$.
The discrete series representations $(\pm,p,j)$ are 
constructed respectively from a highest and a lowest weight
\begin{equation}
  \label{eq:Characters-d}
  \begin{split}
    \chi_{(+,p,j)}(z,u)
    &\ =\ \sum_{n\geq0}z^{j-n}\,u^p
     \ =\ \frac{z^j\,u^p}{1-1/z}\\[2mm]
    \chi_{(-,p,j)}(z,u)
    &\ =\ \sum_{n\geq0}z^{j+n}\,u^{-p}
     \ =\ \frac{z^j\,u^{-p}}{1-z}\  \ .
  \end{split}
\end{equation}
The Casimir operator $C = -\frac{1}{2}(P^+P^-+P^-P^+)-2JK$ assumes
the values $C = p(1 \pm 2j)$ on these representations. 
The continuous series representations $(0,\s,j)$
possess neither a highest nor a lowest weight state and   
the generator $K$ acts trivially on them 
so that their characters are simply given by
\begin{equation}
  \label{eq:Characters-c}
  \chi_{(0,\s,j)}(z, u)
  \ =\ \sum_{n\in\Integer}z^{j + n}\ \ .
\end{equation}
The additional label $\s$ gives the value of the Casimir, $C=s^2$. 

We now review the representation theory of the affine Lie algebra
$\affH$. The simplest class of representations of $\affH$ is generated by
the action of all the negative modes on the unitary representations
of the horizontal subalgebra. These Verma modules turn out to be
irreducible. Let us recall that the Cartan subalgebra of $\affH$
is generated by the current zero modes $J_0$ and $K_0$ and by the
Virasoro generator $L_0$. Following our prescription for the
horizontal subalgebra we thus introduce the character of a
representation $\mu$ as follows,
\begin{equation}
  \label{eq:CharDef}
  \chi_\mu^\affH(q,z,u)\ =\ \tr_\mu\Bigr[q^{L_0-\frac{c}{24}}z^{-iJ_0}u^{-iK_0}\Bigr]\ \ .
\end{equation}
Their explicit form is easily derived 
\begin{align}
  \label{eq:AffineCharactersOne}
  \chi_{(+,p,j)}^\affH(q,z,u)
  &\ =\ \frac{q^{h_{(+,p,j)}-\frac{1}{12}}\,z^j\,u^p}{(1-z^{-1})\eta(q)^2
\prod_{n=1}^\infty(1-zq^n)(1-z^{-1}q^n)} \hspace{1cm} \bigl(\,|q|^{-1}>|z|>1\,\bigr)&\nonumber\\[2mm]
  \chi_{(-,p,j)}^\affH(q,z,u)
  &\ =\ \frac{q^{h_{(-,p,j)}-\frac{1}{12}}\,z^j\,u^{-p}}{(1-z)\eta(q)^2
\prod_{n=1}^\infty(1-zq^n)(1-z^{-1}q^n)} \hspace{1.4cm} \bigl(\,|q|<|z|<1\,\bigr)& \nonumber \\[2mm]
  \chi_{(0,\s,j)}^\affH(q,z,u)
  &\ =\ \frac{q^{h_{(0,\s,j)}-\frac{1}{12}}\,\sum_{n\in\Integer}z^{j+n}}{\eta(q)^2
\prod_{n=1}^\infty(1-zq^n)(1-z^{-1}q^n)}
   \ =\ \frac{q^{h_{(0,\s,j)}}}{\eta(q)^4}\sum_{n\in\Integer}z^{j+n}\ \ .& &  
\end{align}
We will refer to the modules $(\pm,p,j)$ and $(0,\s,j)$ as
standard representations. Furthermore, we restrict the range
of the label $p$ of the  $(\pm,p,j)$ representations to the interval $(0,1)$.
\smallskip

The conformal weights of the ground states transforming in
the representations $(\pm,p,j)$ and $(0,\s,j)$ are given by
\begin{align}
  \label{eq:ConformalDimensions}
  h_{(\pm,p,j)}&\ =\ \frac{p}{2}(1-p)\pm pj\quad,&
  h_{(0,\s,j)}&\ =\ \frac{s^2}{2}\ \ . 
\end{align}
These value agree with the eigenvalue of the ``improved''  Casimir
operator $\frac{1}{2}(C+K^2)$ that appears in 
the quantum corrected energy momentum tensor. 

In addition to the standard representations, $\affH$ also admits
representations which in some loose sense can be referred to as
twisted highest weight representations. Their construction rests
on the observation that the $\affH$ current algebra \eqref{affh4}
allows for a spectral flow automorphism which acts on the modes
as (see \cite{dk} for instance)
\begin{align}
  \label{eq:SF}
  \Omega_\omega(P_n^\pm)&\ =\ P_{n\mp \omega}^\pm&
  \Omega_\omega(J_n)&\ =\ J_n&
  \Omega_\omega(K_n)&\ =\ K_n + i\omega\delta_{n0}\ \ .
\end{align}
From this definition one readily derives the action
\begin{equation}
  \label{eq:SFVir}
  \Omega_\omega(L_n)\ =\ L_n-i\omega J_n
\end{equation}
on the Virasoro modes.
  
Given any representation $\mu$ implemented on a space $\mc{H}_\mu$
via the map $\rho_\mu:\affH\to\End(\mc{H}_\mu)$ one can
define a new representation $\mu_\omega$ which acts on the same
space. The new representation $\mu_\omega$ is implemented through the
concatenation $\rho_{\mu_\omega}=\rho_\mu\circ\Omega_{-\omega}$.
In view of its construction it is termed spectral flow representation.
The previous relations, \eqref{eq:SF} and \eqref{eq:SFVir}, imply a
simple formula for the characters. Indeed, it is straightforward to relate
the character of $\mu_\omega$ to that of the unflowed representation
$\mu$. A simple algebraic manipulation within the trace implies the following
formula for their characters
\begin{equation}
  \label{eq:SFChar}
  \begin{split}
    \chi_{\mu_\omega}^\affH(q,z,u)
     \ =\ u^{-\w}\,\chi_\mu^\affH(q,zq^{-\omega},u)\ \ .
  \end{split}
\end{equation}
In order to simplify the notation we identify the label $\mu_{\omega=0}$ with
$\mu$ whenever there is no chance of confusion.
\smallskip

In contrast to the case of affine Lie algebras based on compact real
forms of finite dimensional semisimple Lie algebras the spectral
flow representations of $\affH$ are not equivalent to the
ordinary highest weight representations introduced before. Instead
as can be seen from formula \eqref{eq:SFChar} they allow to
extend the range of the values of $p$ from the interval $(0,1)$
to the full real axis. 

There is an alternative expression for the characters of the discrete series
which we used repeatedly in our derivation of the coset characters. We can write
\begin{equation}
  \label{eq:AffineCharactersTwo}
  \chi_{(\pm,p,j)_{\mp \w}}^\affH(q,z,u)
  \ =\ u^{\pm (p+\w)} \frac{1}{\eta(q)^4} \sum_{n\in\Integer}z^{j  \mp n}\sum_{m=0}^\infty(-1)^m
q^{h_{(\pm,p,j)} \pm \w(j \mp n) + \frac{m}{2}(m + 2n + 1)}\ \ ,
\end{equation}
where $\w \in \mathbb{N}$.
With the upper sign 
this formula is valid for  $|q|^{-\w}<|z|<|q|^{-1-\w}$
while it is valid for $|q|^{1+\w}<|z|<|q|^{\w}$ if we choose the lower sign.
This expression for the characters follows from the identity
\begin{equation}
  \label{eq:CharIdOne}
  \frac{1}{(1-z)\prod_{n=1}^\infty(1-zq^n)(1-z^{-1}q^n)}
  \ =\ \sum_{n\in\Integer}z^n\sum_{m=0}^\infty(-1)^m\frac{q^{\frac{m}{2}(m+2n+1)+\frac{1}{12}}}{\eta(q)^2}\ \ ,
\end{equation}
where $|q|<|z|<1$ is assumed. 
This relation, 
originally proved in \cite{Pak}, 
can also be derived in the following way  \cite{Rib}.
We first write
\begin{equation}
  \label{eq:FormulaOne}
  \frac{1}{(1-z)\prod_{n=1}^\infty(1-zq^n)(1-z^{-1}q^n)}
  \ =\ \sum_{l\in\Integer}f_l(q)\,z^l\ \ ,
\end{equation}
and then multiply both sides with $z^{-n-1}$ and perform an integral
along a contour contained in the interior of the unit circle.
On the right hand side of \eqref{eq:FormulaOne}
we simply obtain the coefficient $f_n(q)$ while on the left hand side we pick up 
the residues from the poles at $z = q^m$, $m>0$. The result
\begin{equation}
  f_n(q)\ =\ \sum_{m=1}^\infty(-1)^{m-1}\frac{q^{-mn+\frac{1}{2}m(m-1)+\frac{1}{12}}}{\eta(q)^2}
        \ =\ \sum_{m=0}^\infty(-1)^m\frac{q^{\frac{1}{2}(m+1)(m-2n)+\frac{1}{12}}}{\eta(q)^2}\ \ ,
\end{equation}
coincides with \eqref{eq:CharIdOne} after the shift $m\to m+2n$.

\end{document}